\documentstyle[referee]{mn}
\def\etal{{\it et al. }}
\def\ie{{\it i.e.  }}

\title[On X--ray emission from GRBs...]
	{On the anomalous X--ray afterglows of GRB 970508 and GRB 970828}
\author[M. Vietri et al.]
	{ M. Vietri$^{1}$, C. Perola$^{1}$,
	 L. Piro$^2$, and L. Stella$^{3}$\thanks{Affiliated to ICRA} 
\\$1$ Universit\'a di Roma 3, Via della Vasca Navale 84 - 00147 Roma, Italy
\\$2$ Istituto di Astrofisica Spaziale, 00040 Frascati (Roma) Italy
\\$3$ Osservatorio Astronomico di Roma, Via Frascati, 33 - 00040
                Monte Porzio Catone (Roma), Italy
}

\date{}

\begin{document}

\maketitle

\label{firstpage}

\begin{abstract}
Recently, BeppoSAX and ASCA have reported an unusual resurgence of soft X--ray 
emission during the afterglows of GRB 970508 and GRB 970828, together with 
marginal evidence for the existence of Fe--lines in both objects. 
We consider the implications of the existence of a torus of iron--rich material 
surrounding the sites of gamma ray bursts as would be expected in the SupraNova 
model; in particular, we show that the fireball will
quickly hit this torus, and bring it to a temperature $\approx 3\times 10^7
\; ^\circ K$. Bremsstrahlung emission from the heated up torus will
cause a resurgence of the soft X--ray emission with all expected characteristics
(flux level, duration and spectral hardening with time) identical to those 
observed during the reburst. Also, thermal emission from the torus will
account for the observed iron line flux. These events are also observable,
for instance by new missions such as SWIFT, when beaming away from our line 
sight makes us miss the main burst, as Fast (soft) X--ray Transients, with 
durations $\approx 10^3\; s$, and fluences $\approx 10^{-7}-10^{-4}\; erg\; 
cm^{-2}$. This model provides evidence in favor of the SupraNova model 
for Gamma Ray Bursts.
\end{abstract}


\section{Introduction}

The discovery of afterglows from gamma ray bursts has greatly strengthened our 
confidence in the correctness of the fireball model (Rees and M\'esz\'aros 
1992). Since then, attention has begun to shift toward the nature of the 
exploding source, a problem which is conveniently decoupled from the fireball 
itself and the ensuing afterglow. For this reason, evidence about the nature of 
the source has to be sought elsewhere. In particular, attention has been called 
to the possible interaction of the burst with surrounding material, and the 
possible generation of a detectable Fe line in the soft X--rays
(Perna and Loeb, 1998, Boettcher \etal., 1999, Ghisellini \etal, 1999, 
M\'esz\'aros and Rees 1998).

Recently, a reburst, \ie, a resurgence of emission during the afterglow has
been reported in two bursts, GRB 970508 (Piro \etal, 1998), and GRB 970828 
(Yoshida \etal, 1999). In the case of GRB 970508, 
the reburst occurs about $10^5\; s$ after the burst, with the soft X--ray flux 
clearly rising, and departing from its otherwise typical power--law decline. 
This resurgence lasts a total of 
$\approx 4\times10^5\; s$, reaches a typical flux in the BeppoSAX band of 
$\approx 8\times10^{-13} \:erg \;s^{-1}\; cm^{-2}$, after subtraction of the 
normal afterglow, and shows evidence for a harder spectrum than during the 
afterglow proper (power law photon index of $\alpha = 0.4\pm 0.6$, as 
opposed to $\alpha =1.5\pm0.6$ before the reburst, and $\alpha = 2.2\pm0.7$ 
at the end of the reburst), (Piro \etal, 1998, 1999). 

Furthermore, possible evidence for the existence of Fe K-shell emission lines 
has been found in these same two bursts: for GRB 970508 see Piro 
\etal, 1999, while for GRB 970828 see Yoshida \etal 1999. In the first case, a 
$K_\alpha$ iron line occurs at an energy compatible with  
the burst's optically determined redshift, 
while in the second one, for which no independent redshift determination exists,
the line, if interpreted as $K_\alpha$ from neutral, or weakly ionized iron, 
yields a redshift of $z = 0.33$. What is
astonishing are the inferred 
line fluxes and equivalent widths: for GRB 970508, $F = 
(2.8\pm 1.1) \times 10^{-13}\; erg\; s^{-1}\;cm^{-2}$ (EW $\approx 1.1\; keV$), 
while for GRB 970828 $F = (1.5 \pm 0.8) \times 10^{-13}\;erg\;s^{-1}\;cm^{-2}$ 
(EW $\approx 3\; keV$). In the case of GRB 970508, furthermore, 
no evidence for the Fe-line was found after about $10^5\;s$.

Despite their inferred intensities, these lines are at the limit of BeppoSAX 
and ASCA detectability, so that further observations are needed to confirm 
their presence. On the contrary the statistical significance of the rebursts is 
very robust. In the following, we shall concentrate on the especially well
documented case of GRB 970508, keeping in mind that qualitatively similar
arguments apply to GRB 970828 as well.

It is the aim of this paper to show that, if enough material of sufficiently 
high density is present in the surroundings of the gamma ray burst event site,  
then this reburst is exactly what one ought to expect on 
theoretical grounds. In particular, it is possible to explain all 
observed characteristics of the reburst, duration/flux level/spectral
hardening, including the (possible) presence of the iron lines. In the next 
section we shall consider the dynamical interaction of the burst's ejecta with 
the torus, and in the following one we shall discuss the thermodynamic state
of the torus, and establish the properties of its (thermal) emission. In the
discussion, it will also be pointed out that the thermodynamic status of 
the torus is precisely the same {\it postulated} by Lazzati \etal (1999) to 
explain the properties of the iron line.

\section[]{Dynamical interaction with surrounding gas}

Both Piro \etal (1999) and Lazzati \etal (1999) have already argued that the 
material giving rise to the Fe K-line
cannot lie on the line of sight: the ensuing column depths, in H and 
Fe would give effects easy to observe. 
Furthermore, this material should be 
present in large amounts which would spoil the smooth, power--law expansion of 
the afterglow, which is observed to cover more than a year. We thus begin by 
assuming that the site of the explosion is surrounded by a thick torus of 
matter, with an empty symmetry axis pointing roughly toward the observer. The 
particle density $n$ and distance $R$ from the explosion site will be scaled in 
units of $10^{10}\; cm^{-3}$ and $10^{16}\; cm$. 

A time $R/c$ after the explosion, this torus will be inundated by the burst 
proper, and a few seconds later ($\delta\!t \approx R/\gamma^2 c = 30\;s$, 
where $\gamma = 100$ is the shell bulk Lorenz factor), it will be hit by the 
ejecta shell. This crash will generate a forward shock propagating into the 
torus, and a reverse one moving into the relativistic shell. For any reasonable 
value of the torus density, the forward shock will quickly rake up as much mass 
as there is in the shell: we find that this occurs after the shock has 
propagated a mere distance $d$, with
\begin{equation}
d = 6\times10^8\; cm \frac{E}{10^{51}\;erg} \frac{10^{10}\; cm^{-3}}{n}
\left(\frac{10^{16}\;cm}{R}\right)^2 \frac{100}{\gamma}\;.
\end{equation}
As is well--known, this means that the relativistic shell must slow down to 
sub--relativistic speeds. Thus, after just $d/c \approx 0.1\;s$, the forward 
shock has become sub--relativistic. The large pressure behind the forward shock 
acts to steepen the reverse shock, which will thus slow down the incoming 
material to sub--relativistic speeds as well. All of this occurs a few seconds 
after the torus sees the burst. 

The total energy released is expected to be of order of the whole kinetic 
energy of the shell, because post--shock acceleration of electrons occurs at 
the expense of the shell bulk expansion, in the shocks. If we suppose that the 
burst generated a total energy release of $E = 10^{51}\;erg$, that the initial 
burst is roughly isotropic, and that the torus covers $\delta\!\Omega$ radians 
as seen from the explosion site, the total energy release $E_{sh}$ will be
\begin{equation}
E_{sh} = \frac{\delta\!\Omega}{4\pi} E \;.
\end{equation}

The total emission timescale can also be reliably computed: the reader will 
have already noticed that this emission scenario is similar to the external 
shock scenario (M\'esz\'aros, Laguna and Rees 1993), except for two differences.
First, in the external shock scenario we are seeing the burst from a 
reference frame which is moving with respect to the shell of shocked gas
with large Lorenz factor, while here the observer is sitting in a 
reference frame in which the shocked gas is moving sub--relativistically. The 
major consequence of this first difference is that the photon emission will
be isotropic, and we shall thus see it, even though the initial shell movement
was perpendicular to the line of sight. The second difference
is that, in the external shock scenario, it is matter ahead of the forward 
shock which is moving relativistically with respect to the shocked gas,
while matter entering 
the reverse shock is moving only barely relativistically
with respect to it. In this paper, instead, the opposite applies: matter 
entering the reverse shock is relativistic, while the forward shock is
barely, if at all, relativistic. 

Still, these two differences do not spoil the fact that electrons accelerated 
at either shock cool much faster than the shell light--crossing time, as
shown by M\`esz\`aros, Laguna and Rees (1993), so that the total burst 
duration is given by the time the reverse shock takes to cross the whole shell. 
In our model, the shell thickness in the laboratory frame is $\approx R/\gamma$
(M\`esz\`aros, Laguna and Rees 1993), and, since the reverse shock is
relativistic with respect to the incoming matter, the shock crossing time, 
and thus also the duration $t_{sec}$ of the secondary burst, is given by
\begin{equation}
\label{dur}
t_{sec} = \frac{R}{\gamma c} = 3\times10^3\; s \frac{R}{10^{16}\;cm}\;.
\end{equation}
Together, the total energy release and emission timescale give us the expected 
bolometric luminosity; the observed flux can be computed, for cosmological 
parameters $\Omega = 1$, $H_0 = 65\;km\;s^{-1}\; Mpc^{-1}$, and $\Lambda = 0$, 
and knowing the burst's redshift $z = 0.835$ (Metzger \etal, 1997), and is
\begin{equation}
\label{flux}
F_X = 1.5\times10^{-10} erg\;s^{-1}\;cm^{-2} 
\frac{\delta\!\Omega}{4\pi}
\frac{E}{10^{51}\;erg} \frac{10^{16}\;cm}{R}\;.
\end{equation}

We must now establish in which band this emission will end up. As is 
well--known, bursts' spectra are highly variable, both from burst to burst and 
within the same burst, at different moments. Also, the fireball model is not 
too specific about the spectral characteristics of bursts. We can still get an 
idea of the spectrum, however, by noticing first that the spectrum will be 
non--thermal, with the usual power law dependence upon photon energy typical of 
synchrotron emission, and second that once again we are observing a burst in 
the external shock scenario, but in the shell frame. In normal bursts,
the spectrum has a break at an energy $\epsilon_b$, which is approximately 
$\epsilon_b \approx 1\; MeV$. However, this spectral feature 
is blueshifted in the observer's frame by the shell's bulk Lorenz factor: 
$\epsilon_b = \gamma \epsilon_i$. The intrinsic spectral break $\epsilon_i$, 
\ie in the shell frame, is thus given by
\begin{equation}
\label{break}
\epsilon_i = \frac{\epsilon_b}{\gamma} = 10\;keV \frac{\epsilon_b}{1\;MeV}
\frac{100}{\gamma}\;.
\end{equation}

It is clear why this secondary burst was not observed. First of all,
it is dimmer than the original one by a bolometric factor of $\delta\!\Omega/
4 \gamma\pi < 10^{-2}$, which would push it below detection 
threshold for both BATSE and the GRBM/WFC instruments of BeppoSAX. Also,
it must have occurred sometime between the burst proper and the BeppoSAX 
detection of the iron line, when, however, BeppoSAX was not observing with 
its (more sensitive) Narrow Field Instruments. 

The further evolution of the shocked shell is as follows. The material 
that passed through the reverse shock will have an internal energy density
higher than the pre--shock one by a factor $\Gamma^2$, where $\Gamma \approx
\gamma$ is the Lorenz factor of the reverse shock, as seen by the pre--shocked
ejecta shell. For reasonable radiative efficiencies, the post--shocked
matter will have a relativistic velocity dispersion even after the secondary
burst; then, a rarefaction wave will make it expand at the sound speed 
$\approx c/\sqrt{3}$ back into the cavity from which it came. Thus pressure 
behind the forward shock will be reduced on a time--scale $\approx \delta\!R/c$,
where we can again take for the post--reverse shock shell thickness, as an 
order of magnitude, $\delta\!R \approx R/\gamma$. Thus the heated gas expansion 
time--scale is again $\approx R/\gamma c\approx 3\times 10^3 \; s$. 

As the pressure from the post--reverse shock material is reduced, the forward 
shock keeps propagating because of momentum conservation. However, even this 
shock cannot last long, because of the strong counterpressure applied by the 
pre--shock torus. We shall show in the next section that this material will be 
brought up to $T_f \approx 10^8 \; ^\circ K$ by heating/cooling from the 
primary and secondary bursts. Then it can easily be checked that $\rho_s 
c^2 \approx m_p n v^2$, where $\rho_s$, the shell baryon density, is given by 
spreading the total fireball baryon mass, $E/\gamma c^2$, over the shell 
volume, $4\pi R^3/\gamma$, and the torus' velocity dispersion $v$ is purely 
thermal: $v^2 = k T_f/m_p$. Thus the torus counterpressure will halt the
forward shock as soon as it becomes subrelativistic.

We now make a small detour to discuss an interesting point about the 
kinematics. As seen from the observer, the part of the 
shell moving toward him will have moved a long distance ($\approx R$, taking 
the torus to be perpendicular to the line of sight) toward him before the torus 
is reached by the burst, and thus starts emitting. At that point, photons start 
travelling away from the torus, and they will catch up with the part of the 
expanding matter shell moving toward the observer at a rate
\begin{equation}
\delta\!R = (c-v) \delta\!t
\end{equation}
where $v \approx c(1-1/2\gamma^2)$ is the matter speed. However, the time 
appearing in the above equation is the time in the reference frame of the 
exploding object, which is related to that of the observer, $t_o$, by 
$\delta\!t_o = \delta\!t (1-v/c)$, and thus, the distance by which the photon 
catches up with the matter shell, in an observer's time interval $\delta\!t_o$ 
is
\begin{equation}
\delta\!R = c \delta\!t_o
\end{equation}
which is identical to the expression when relativistic effects are not present. 
This immediately allows us to estimate the distance of the torus: in fact, 
since the reburst was present in the observations made $\approx 10^5
\; s$ after the burst, and this can only occur after the bursts' photons have 
reached the torus, we deduce that $R(1-\cos\theta) < 3\times 10^{15}\; cm$,
where $R$ is the torus distance from the line of sight, and $\theta$ is
the angle away from the line sight of the torus symmetry plane. 
For the total distance, we shall take $R \approx 10^{16}\; cm$. 

\section{Thermal history of the torus}

In order to proceed, we need first to determine the torus thickness, which
we do by using a constraint from the observations of the iron line. 
When the torus is reached by the burst proper, the ionization parameter is
\begin{equation}
\xi \equiv\frac{L}{n R^2} = 10^9 
\frac{L}{10^{51}\;erg\;s^{-1}}
\frac{10^{10}\;cm^{-3}}{n}
\left(\frac{10^{16}\;cm}{R}\right)^2\;.
\end{equation}
For these large values, we expect that all iron will be completely ionized, 
so that the generation of the iron line by fluorescence is unlikely. 
Furthermore, the torus will be hit by the secondary burst only $R/\gamma^2 c 
\approx 30\; s$ later: thus fluorescence with afterglow photons cannot be
invoked either. The 
remaining mechanisms, multiple recombination/ionizations and thermal 
processes, both require the torus Thomson optical depth $\tau_T \approx 1$
for maximum efficiency, and to avoid line smearing (fluorescence, instead, 
requires $\tau_T \gg 1$). In such a thin shell, the torus temperature is 
quickly brought up by the primary burst photons to a temperature close to its 
Inverse Compton value, given by $4kT_{IC} = \bar{\epsilon}$, with 
$\bar{\epsilon}$ the average burst photon energy. Taking this to be of order 
the break photon energy $\epsilon_b \approx 1\; MeV$, we find
$T \approx \frac{\epsilon_b}{4k} \approx 3\times 10^9\; ^\circ K$.
However, at this temperature, pair creation will quickly give $\tau_T \gg 1$,
and the ensuing thermal cooling will badly limit the temperature, to a value 
close to the pair creation limit, 
\begin{equation}
\label{tic}
T_{IC} \approx 5\times 10^8\; ^\circ K\;.
\end{equation}
At such large temperatures, the bremsstrahlung cooling time--scale is
quite long $t_{br} \approx 5\times 10^5 s (10^{10}\;cm^{-3}/n) (T/ 5\times 
10^8\;^\circ K)^{1/2}$. However, the torus may cool due to Inverse Compton
cooling off the photons produced by the crashing of the ejecta onto the
torus, which have a typical photon energy $\epsilon_i$ (Eq. \ref{break}) 
much below the torus temperature. For ease of reference, we shall call these
secondary photons. The Inverse Compton cooling time--scale
$t_{IC} = 3 m_e c^2/8 c\sigma_T U_{ph}$ (where $m_e$ is the electron's
mass, and $\sigma_T$ the Thomson cross--section), can be computed using the
fact that the photon energy density $U_{ph} = L/c A$, where $L$, the
secondary photons' luminosity, was given above as $L = 
E \delta\!\Omega/4\pi t_{sec}$, and the total area is roughly twice the 
shock area, $ A \approx 2 R^2 \delta\!\Omega$. We find thus $U_{ph} = 
E\gamma/8\pi R^3$, independent of the solid angle subtended by the torus.
The ratio of the Inverse Compton cooling time to the duration of the
secondary burst is then given by
\begin{equation}
\label{ratio}
\frac{t_{IC}}{t_{sec}} = \frac{3\pi m_e c^2 R^2}{\sigma_T E} = 
1.3 \left(\frac{R}{10^{16}\; cm} \right)^2 \frac{10^{51}\;erg}{E}\;.
\end{equation}
We see that this ratio is very sensitive to the torus location, and to the
total energetics. For $t_{IC} \geq t_{sec}$, the torus matter will 
remain hot (Eq. \ref{tic}), while for $t_{IC} <  t_{sec}$, its 
temperature will cool to the new Inverse Compton temperature of the
secondary photon bath:
\begin{equation}
T^{(2)}_{IC} \approx \frac{\epsilon_i}{4k} \approx 3\times 10^7\; ^\circ K\;.
\end{equation}
For the parameters assumed here, $t_{IC}\approx t_{sec}$, so that the torus 
will probably settle to a value intermediate between $T^{(2)}_{IC}$ and
$T_{IC}$. We scale the value of $T$ to $T_f = 10^8\;^\circ K$, but see 
the next section for a discussion. 

The bremsstrahlung cooling time, at this lower temperature, is given by 
$t_{br} \approx 1.3\times 10^5 \;s (10^{10}\;cm^{-3}/n)$, comparable to the
total duration of the reburst observed by Piro \etal, 1998. Also, the 
expected flux level is
\begin{equation}
\label{fluxbrems}
F_{br} = 1.1\times 10^{-12} \; erg\; s^{-1}\; cm^{-2} 
\left(\frac{M}{1 M_\odot}\right)^2  \frac{10^{46} cm^3}{V} 
\left(\frac{T}{10^8\;^\circ K}\right)^{1/2}\;,
\end{equation}
provided the torus cooling time is longer than the torus light crossing time,
$t_{lc} \approx R/c$. Otherwise, the observed flux $F_{br}^{(obs)}$ would be 
related to the previous formula by
\begin{equation}
F_{br}^{(obs)} = F_{br}\times \frac{t_{br}}{t_{lc}}
\end{equation}
Furthermore, when, initially, the temperature is rather large,
$\approx 10^8\; ^\circ K$, the spectral slope 
between the BeppoSAX' Low and Medium Energy concentrator optics/spectrometers
should be rather flat, while later, as the torus cools and its flux decreases,
the spectral slope should also increase. Piro \etal (1999) find that, at the
point where the reburst is (fractionally) highest over the smooth afterglow, 
$\alpha = 0.4\pm 0.6$ (\ie, consistent with a flat 
bremsstrahlung spectrum), while later they find $\alpha = 2.2\pm 0.7$. 
Though there are large errors, the steepening of the spetrum through the reburst
appears to be significant. In view of the agreement of the duration timescale, 
flux level, and steepening of spectral slope, we suggest that the observed
reburst in GRB 970508 is thus bremsstrahlung radiation from a torus of
hot material, heated up, and then cooled down, by the photons produced by the
impact of the burst ejecta.

We now need to cover our tracks by determining whether there are values of
the total torus mass and volume which satisfy, together with $F_{br}^{(obs)} =
1\times10^{-11}\; erg\;s^{-1}\;cm^{-2}$, also $\tau_T \approx 1$, $n 
\approx 10^{10}\;cm^{-3}$ which we assumed throughout. We assume a geometry
whereby the torus has a volume $V = \delta\!\Omega R^2 \;\delta\!R$, with
the torus thickness $\delta\!R \leq R$, the torus distance from the explosion 
site. Since $\tau_T = 0.6 (M/1 M_\odot) (10^{16}\;cm/R)^2 4\pi/\delta\!\Omega$, 
we see that for $M = 5 M_\odot$, $R = 10^{16}\; cm$, $V = 10^{47} cm^3$,  
$\delta\!R = 10^{14}\; cm $ and $\delta\!\Omega \approx 4\pi$, we satisfy all 
constraints simultaneously: $\tau_T \approx 2$ and $n = 4\times 10^{10}\;
cm^{-3}$. From this we see that the torus need not be thin ($\delta\!\Omega
\approx 4\pi$), which certainly agrees with expectations about the
nature of exploding sources. Also, we notice that $t_{br}/t_{lc} \approx 4$, 
so that the duration of the bremsstrahlung cooling radiation is diluted 
by light crossing time effects. 

Thermal expansion of the shell during the cooling phase is
negligible, since the cooling time is of order of the light crossing time,
which is certainly shorter than the sound crossing time. 

It is well--known that GRB 970508 had an early optical detection, $\approx 
0.2^d$ after the burst, which was dimmer than later ($ > 1^d$) detections
(Sahu \etal, 1998). Typical fluxes throughout the first 2 days are around $30\; 
\mu Jy$, which far exceed the optical component of the bremsstrahlung emission 
from the torus, which is in the range of $\approx 0.03\; \mu Jy$. Thus the
observed nearly simultaneous rise of X--ray and optical fluxes remains, within 
this model, a coincidence. 

\section{Discussion}

Beyond explaining the observed X--ray reburst (and the Fe line, see below), 
the current model makes a number of interesting predictions. First, the
secondary burst may be observable. We may expect these events to last a few 
thousand seconds, with fluxes in the range of $10^{-11}$ to $10^{-10}\; erg\; 
s^{-1} \; cm^{-2}$. The spectra of these sources are also interesting: we
argued above that the torus temperature is limited by pair--creation, which
would otherwise cause excessive radiative losses; thus we may expect the
torus to reach a limiting temperature such that $\tau_T \approx 1$, and
a temperature $5\times 10^8 \; ^\circ K$, which correspond to a Compton 
parameter $y \approx 0.5$. We thus expect significant departures from the
usual, power--law like spectra of bursts. In particular, from sources 
which do not have time to cool down to $T_f$ (Eq. \ref{ratio}), so that the 
Comptonization of the secondary burst spectrum is time--independent, we 
expect to see a cutoff $\propto \exp(-h\nu/kT)$ beyond $h\nu = kT
\approx 50\; keV$, with a complicated, time--dependent non--power--law 
behavior below this point (Rybicki and Lightman 1979).
This exponential cutoff can be used as signature of unsaturated 
Comptonization, typical of the present model. 

Another interesting consequence of this model is that the secondary burst
may be seen even without its being preceded by the main gamma ray burst. This
would occur whenever we would missed the (beamed) emission from the burst
proper, but would see the isotropic emission from the reburst. This might
occur because in many models, the beaming of the main burst is expected
to be rather smooth, and one may conjecture that, while the total output
may be $\approx 10^{52}\; erg$ close to the major axis, a total of $10^{51}\;
erg$ remains to be emitted nearly isotropically. This would amply satisfy the
energy requirements of the reburst. The total expected fluences (up to 
$\approx 10^{-4} \; erg\; cm^{-2}$ for distances smaller than GRB 
970508's)) and durations ($\approx 10^3\;s$) strongly remind one
of the so--called Fast X--ray Transients, many of which last through several
satellite orbits and have no identified counterparts (Grindlay 1999).
A bevvy of these events should become observable with planned new telescopes 
such as SWIFT. 

An interesting question one may ask is why the observation of the rebursts
is so rare: up to now, GRB 970508 and GRB 970828 are the only two bursts
for which such phenomenon has been observed. So long as the torus is optically
thin to bremsstrahlung, we see from Eq. \ref{fluxbrems}
that the expected flux scales with distance from the explosion site as
$R^{-q}$, where $q = 2-3$. Since we ignore the torus thickness, we consider
the two limiting cases: $q=3$, uniformly filled sphere, and $q=2$, infinitely
thin shell. This flux will appear with a time--delay $R/c$ with respect to
the burst, simultaneously with an afterglow which scales as $t^{-p}$, with
$p \approx 1.3$. We see that the torus to afterglow flux ratio scales as
$t^{p-q}$, with $p-q = -(0.7-1.7) < 0$. Thus, the more distant the torus
is, the less easy it is to detect it. However, since we supposed that 
$\tau_T \approx 1$ for $R\approx 10^{16}\; cm$, further shrinking of the 
torus will make it less bright, not more; but it will have to compete with
a simultaneously emitted afterglow which is brighter and brighter. So
$R = 10^{16}\; cm$ is an ideal distance at which the torus could be located. 

For the same parameters as above, Lazzati \etal (1999) have shown that the 
iron line can be interpreted as due to purely thermal processes. Actually,
Lazzati \etal showed that also fluorescence 
and multiple ionization/recombinations
can account for the line, given suitable (but different!) thermodynamic
conditions for the emitting plasma. However, we showed in this paper
that the thermodynamic conditions of the emitting torus are not free, but
are essentially fixed by the requirement that the reburst be fitted. We wish
to stress that this is a much more demanding requirement, since the reality
of the reburst cannot be doubted, while that of the iron line is more 
questionable. It is however satisfying that the thermodynamic parameters
thusly determined ($T = 10^8\; ^\circ K$, $n = 4\times 10^{10}\; 
cm^{-3}$) are precisely those that Lazzati \etal (1999) had to assume,
in order to fit the line. 

As a corollary, one may then understand why it is difficult to observe 
the iron lines. Lazzati et al. (1999) have derived the luminosity of the 
line as a function of the torus temperature: $\propto \exp(-8\times 10^7\; 
^\circ K/T) T^{-2.4}$. This luminosity has a peak for $T = T_m = 3\times 10^7 \;
^\circ K$, and decreases steeply with increasing $T$. We see that $T^{(2)}_{IC} 
\approx T_m$, while $T_{IC} \gg T_m$. Thus, it is only 
when the torus manages to cool down, that it will find itself in ideal 
conditions for producing a bright iron line; we see from Eq. \ref{ratio}
that this occurs only for material that lies close to the explosion site.
Otherwise, the torus material will remain into 
a hot state in which the line equivalent width is very small: $\approx 20\; eV$ 
at $T = T_{IC}$ (Bahcall and Sarazin 1978). We also remark that, even in the
case in which the torus has managed to cool down to $T_m$, after a time 
$t_{br}$, it will further cool below $T_m$, and the line flux will promptly 
decrease, thereby explaining the disappearance of the iron line in the 
observations of GRB 970508 (Piro \etal, 1999). 

Should the torus be located at larger radii, then we would expect that
the material be hotter (from Eq. \ref{ratio}), and that the Fe--line 
should not be observable, from the argument above. We thus expect inverse
correlations of the time--delay with which the reburst appears with the 
luminosity, and with the Fe--line equivalent width.

An alternative model for the anomalous behaviour of GRB 970508 has been
proposed (Panaitescu, M\`esz\`aros and Rees 1998). In their model there is
no external material to cause a resurgence of the X--ray flux, and the 
peculiarities in the time--evolution of the optical afterglow are explained 
as a consequence of beaming. However, the anomalous 
variations in the X--ray flux can hardly be followed (see especially their 
Fig. 2), and certainly there is no allowance for either the observed spectral 
variations of the X--ray flux during the first two days, nor for the existence 
of an iron line. 

Lastly, we would like to comment on the fact that we require a dense, and 
abundant amount of iron--rich (for a redshift of 
$z=0.835$!) material, at close distance from
the explosion site: $5 M_\odot$ at $R = 10^{16}\;cm$. This is 
clearly incompatible with all existing models of GRBs, neutron star--neutron 
star/neutron star--black hole/ black hole--white dwarf mergers, and hypernovae, 
except for SupraNovae (Vietri and Stella 1998), which are preceded by a 
SuperNova explosion occurring between 1 month and 10 years before the GRB. With 
an average expansion speed of $3000\; km\;s^{-1} $, this implies an accumulated 
distance of $R = 10^{15-17}\; cm$. At this distance, one should find 
several solar masses (McCray 1993) with densities of order $10^{10}\; cm^{-3}$, 
exactly as required by this independent set of observations.

\bsp

\label{lastpage}

\end{document}